\documentstyle[12pt,epsfig,cite]{article}
\newcommand{\beq}{\begin{equation}}
\newcommand{\eeq}{\end{equation}}
\newcommand{\bea}{\begin{eqnarray}}
\newcommand{\eea}{\end{eqnarray}}

\newcommand{\gsim}{\lower.7ex\hbox{$
\;\stackrel{\textstyle>}{\sim}\;$}}
\newcommand{\lsim}{\lower.7ex\hbox{$
\;\stackrel{\textstyle<}{\sim}\;$}}

\def\cp{{\bf CP}}

\begin{document}
\thispagestyle{empty}
\vspace*{-22mm}

\begin{flushright}
UND-HEP-06-BIG\hspace*{.08em}03\\
hep-ph/0603087\\

\end{flushright}
\vspace*{1.3mm}

\begin{center}
{\LARGE{\bf
THE BATTLE OF ALBUERA, THE FC \vspace*{1mm} \\ 
LIVERPOOL AND THE STANDARD \vspace*{3mm} \\  
MODEL
}}
\vspace*{19mm}

{\Large{\bf I.I.~Bigi}} \\
\vspace{7mm}

{\sl Department of Physics, University of Notre Dame du Lac}
\vspace*{-.8mm}\\
{\sl Notre Dame, IN 46556, USA}\\
{\sl email: ibigi@nd.edu}
\vspace*{10mm}

{\bf Abstract}\vspace*{-1.5mm}\\
\end{center}

\noindent 
The Standard Model despite its well-known short comings is unlikely to yield without offering stubborn resistance. There are compelling arguments that New Physics lurks around the TeV scale. 
Continuing comprehensive studies of beauty, $\tau$ and charm transitions can be instrumentalized 
to reveal it and shed light on it. They are thus complementary to findings obtained at the LHC and presumably essential in clarifying the true nature of that New Physics. A Super-B facility seems to provide the cleanest environment for pursuing such an ambitious program. I list desirable features of such a setup as well as challenges for the accelerator and detector designs and for the theoretical analysis. 

\section{The Verdict on the SM}
\label{VERDICT}

After losing the battle of Albuera in 1811 Marechal Soult declared: "I had beaten the British -- 
it was just they did not know when they were beaten." 
Soult was actually one of Napoleon's best generals, and experts agree he was right on both 
counts. For those with a much shorter memory span one can point to a similar experience just last year: 
at halftime in the finals of the European Champions League AC Milano was 
leading FC Liverpool 3:0 with truly gorgeous play, yet the pesky Brits while still 
being outplayed in the second half -- except for those magic eight minutes -- refused to concede.  

This is the story as well with the Standard Model (SM): We all know how to design an extension to the 
SM that is greatly superior to it -- now we have to overcome the SM's refusal to concede defeat. 

Even in the last few years since the turn of the millenium the SM has scored unprecedented successes 
in flavour physics: CKM dynamics describe a vast array of very diverse phenomena 
culminating in \cp~violation as observed in particle decays as CKM's signature achievement. Yet we 
are in search of a `New \cp~Paradigm': for we know that CKM dynamics is grossly inadequate 
for baryogenesis, which posits the observed baryon number of the Universe as a 
{\em dynamically generated} quantity rather than an {\em initial input value}. There are further shortcomings of the SM as revealed mostly by heavenly data:  (a) $\nu$ oscillations; 
(b) `dark matter'; 
(c) `dark energy'. 

In addition there are serious explanatory deficits of a general nature (I am even not including the 
so far unresolved `Strong \cp~Problem'): 

{\bf (i)} 
{\em Electroweak Symmetry Breaking and the Gauge Hierarchy}: What are the dynamics driving the electroweak symmetry breaking of 
$SU(2)_L\times U(1) \to U(1)_{QED}$? How can we tame  the instability of Higgs dynamics with its quadratic mass divergence? 
I find the arguments compelling that point to New Physics at the 
$\sim 1$ TeV scale -- like low-energy SUSY; therefore I call it the `confidently predicted' New Physics 
or {\bf cpNP}. 

{\bf (ii)} 
{\em Quantization of electric charge}: While electric charge quantization 
$Q_e = 3 Q_d = - \frac{3}{2} Q_u$ 
is an essential ingredient of the SM -- it allows to vitiate the Adler-Bell-Jackiw or triangle anomaly -- it does not offer any 
understanding. It would naturally be explained through Grand Unification at very high energy scales 
implemented through, e.g., $SO(10)$ gauge dynamics, where leptons and quarks are placed in the same multiplet. 
I call this the `guaranteed New Physics' or {\bf gNP}. 

{\bf (iii)} 
{\em Family Replication and CKM Structure}: We infer from the observed width of $Z^0$ decays that there are  three (light) neutrino species. The hierarchical pattern of CKM parameters as revealed by the data is so peculiar as to suggest that some other dynamical 
layer has to underlie it. I refer to it as `strongly suspected New Physics' or {\bf ssNP}. 
We are quite in the dark about its relevant scales. 
Saying we pin our hopes for explaining the family replication on Super-String or M theory is a scholarly way of saying 
we have hardly a clue what that {\bf ssNP} is.

\section{On Finding What Drives the Electroweak Symmetry Breaking}
\label{FIND}

The next big challenge to which we have to rise is to find {\em and identify} the {\bf cpNP}. It has 
provided the justification for the LHC and drives the motivation for the ILC -- an excellent 
one in my view. 

Let me make two judgment calls. While I have reflected on them, I understand that reasonable 
people can honourably disagree. 
\begin{itemize}
\item 
Any future facility has to be justified by its ability {\em to find New Physics and identify its salient 
features} -- learning new lessons on QCD will no longer suffice. This applies also to a new 
$\tau$-charm factory {\em beyond} BESIII. 
\item 
Heavy flavour studies {\em might} provide insights into questions {\bf (ii)} \& {\bf (iii)} listed above -- 
but we cannot {\em count} on it. Therefore we can{\em not} justify a new facility with such a hope. 
\end{itemize}

Instead I advocate {\em instrumentalizing} studies of flavour dynamics as expressed below 
through five statements: 
\begin{enumerate}
\item 
{\em Comprehensive and detailed} heavy flavour studies will be {\em crucial in identifying} the 
{\bf cpNP}.  
\item 
I remain skeptical that studies in hadroproduction can be fully competitive with those at $e^+e^-$ 
machines in $\tau$, charm and even beauty transitions as far as  
precision and comprehensiveness are concerned. 

In this context I want to emphasize that $B_d$ and $B_s$ decays represent truly different, 
yet complementary chapters in `Nature's Book on Fundamental Dynamics'. 
\item 
A Super-B factory allows comprehensive precision studies of $B$, $\tau$ and charm decays. 
A very detailed plan for such a project has been developed by the KEKB team \cite{SUPERKEK}. 
\item 
To be competitive in $\tau$ \& charm studies a {\em future} $\tau$-charm factory has to be of the 
Super-$\tau$/charm variety, i.e.  with a luminosity of at least $10^{34}cm^{-2}s^{-1}$. 
\item 
I am convinced that a compelling justification for a 
Super-B facility can be given -- yet one cannot merely follow the lines of argument given 
originally in favour of a $B$ factory. There 
\begin{itemize}
\item 
one had so-called `killer-applications', namely \cp~violation in 
$B_d \to \psi K_S$, $\pi^+\pi^-$ and  $K\pi$ 
\item 
with predictions of reasonable accuracy 
\item 
requiring a luminosity in the $(10^{33}-10^{34})cm^{-2}s^{-1}$ range. 

\end{itemize}
While history repeats itself, it never does so in an identical fashion. 
For a Super-B project we face a fundamentally different `landscape': 
\begin{itemize}
\item 
We can{\em not count} on killer applications. 
\item 
We can{\em not count} on a {\em numerically} massive intervention by New Physics. 
\item 
`Merely' finding New Physics is {\em not} enough -- we must identify its salient features. 
\item 
There is no clear benchmark for the needed luminosity. 
\item 
Thus our guidance has to come from what -- rather unkindly -- has 
been referred to as the `Wall Street 
mantra of greed': "Lots is good, more is better, aim for the sky!"
\end{itemize}
\end{enumerate}

There are basically two kinds of research: 
\begin{itemize}
\item 
One has a more or less well-developed theoretical framework, where one has at least clarified the categories 
of relevant questions. Answering those (with the help of experiment) can be called 
`{\em hypothesis-driven}' research. In that case one has always something to show for one's efforts. Not 
surprisingly such projects are most popular with funding agencies. 
\item 
Alternatively one has a situation that is unsatisfactory in a conceptual or even phenomenological way, 
yet with no compelling theory candidate to fill the gap. Without such guidance one 
performs `{\em hypothesis-generating}' research in the hope that more analyses will point to a new 
paradigm. Such work thus has the potential to lead to a revolution -- alas funding agencies display markedly less enthusiasm for it. 

\end{itemize}

The program at the $B$ factories has {\em primarily} been of the {\em hypothesis-driven} variety -- 
and a most successful one at that. Yet at a Super-B factory (with $\tau$ and charm) we have to 
conduct  {\em hypothesis-generating} research with one of the goals being the search 
for the `New \cp~Paradigm'. 

\section{Challenges for a Super-B Facility}
\label{CHALL}

Precision in acquiring and interpreting data is essential if we want to draw the 
desired lessons from heavy flavour studies. The `conditio sine qua non' is to have huge statistics 
of comprehensive high quality data. A large body of well measured transitions is more important than a few rates determined with infinite precision. 

There is the (in)famous challenge from Sanda: "We need a luminosity of $10^{43}cm^{-2}s^{-1}$!" 
While it is certainly `tongue-in-cheek', it is not just frivolous; it has more than a kernel of truth, in particular when combined by Sanda's empirical conjecture that every second `3 sigma' effect goes 
away. 

Yet,  as already stated, statistics is not all. For the goal of a Super-B facility has to go beyond `doing 
more of the same'. We need not only more data, but also data of a different and higher quality. 
New observables have to be opened up to detailed study. This requires a hermetic detector operating 
in a low background environment with superb $\mu$vertex resolution. This would allow to study transitions like $B\to \tau \nu$, $\tau^+\tau^-$l, $\tau \nu X$, $\nu \bar \nu X$ and 
$B\to \gamma X_s$ vs. $\gamma X_d$. 

Also energy flexibility would be very desirable, i.e. to study $\Upsilon (5S) \to B_s \bar B_s$ and even 
$\psi (3770) \to D \bar D$ etc. in addition to $\Upsilon (4S) \to B \bar B$. 

Finally it would be quite desirable to have a {\em polarized} electron beam available. It would lead 
to the production of polarized $\tau$ leptons and probably also of polarized charm baryons. This 
polarization would be a powerful tool to enhance the sensitivity to \cp~violation in the decays of those states; at the same time it would help to control systematics.  

\begin{figure}[t]
\vspace{6.0cm}
\includegraphics{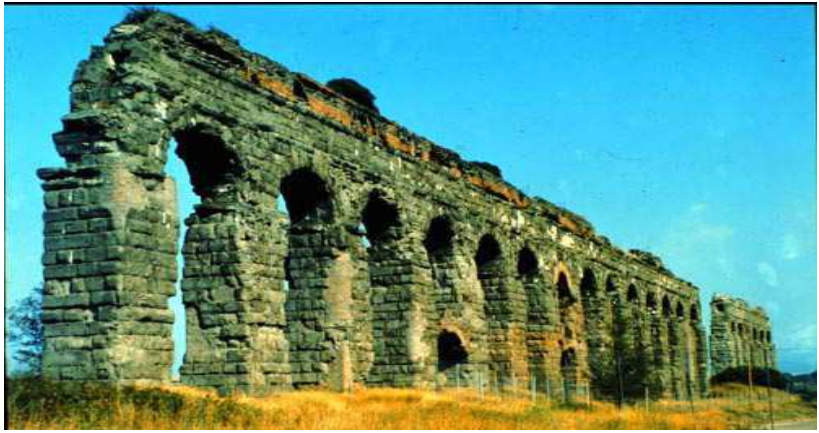}
 \caption{\it
      A prominent linear machine near Rome.
    \label{AQUA} }
\end{figure}

The area around Rome has an ancient history of superbly engineered and long lasting linear structures, see Fig.\ref{AQUA}. The stated aim for an `ILC inspired' Super-B facility 
\cite{ILCSUPERB} is to achieve a luminosity of 
$10^{36}cm^{-2}s^{-1}$ (or more) with tiny beams and a hermetic detector, maybe even with a polarized beam, `soon' and `here', i.e. near Rome. 

Life teaches us all too often that if something is too good to be true -- it usually is. Is it in this case? 
Keep in mind we cannot afford failure. 

\section{Questions and Challenges}
\label{OUT}

Let me pose to you some questions that I would like to see addressed -- or better still answered -- at 
the workshop or in the near future. 
\begin{itemize}
\item 
What integrated luminosity can be achieved at a Super-B factory by 2016, i.e after mature data 
taking has taken place at the LHC, and by 2020, when a realistic optimist can hope for the ILC to begin running?
\item 
How and when can the feasibility of the linear Super-B concept be established? 
\item 
What will be the quality of the $e^{\pm}$ beams, and how hermetic can the detector be? 
\item 
What kind of integrated luminosity can be achieved in, say, a two year run at the 
$\Upsilon (5S)$ run? 
\item 
How many {\em precision} measurements can be made by (an upgraded) LHCb?
\item 
Can (an upgraded) LHCb do competitive \cp~studies in charm transitions?
\item 
How feasible is a Super-$\tau$-Charm factory with $(10^{34}-10^{35})cm^{-2}s^{-1}$, and 
what is its price tag? Would it be competitive with respect to \cp~searches in $\tau$ decays? 
\item 
While the items listed so far mainly concern experimental and technical issues, there is a lot to be done by interested theorists as well: to identify the features of the conjectured New Physics one has to state 
the required 
\begin{itemize}
\item 
benchmark observables \footnote{I realize that considerable work has already been done in this 
direction.}; 
\item 
benchmark accuracy and 
\item 
validation checks for establishing control over theoretical uncertainties. 
\end{itemize}
It is also crucial to interpret findings from the LHC -- including `no-shows'. 

\end{itemize}
\begin{figure}[t]
\vspace{6.0cm}
\includegraphics{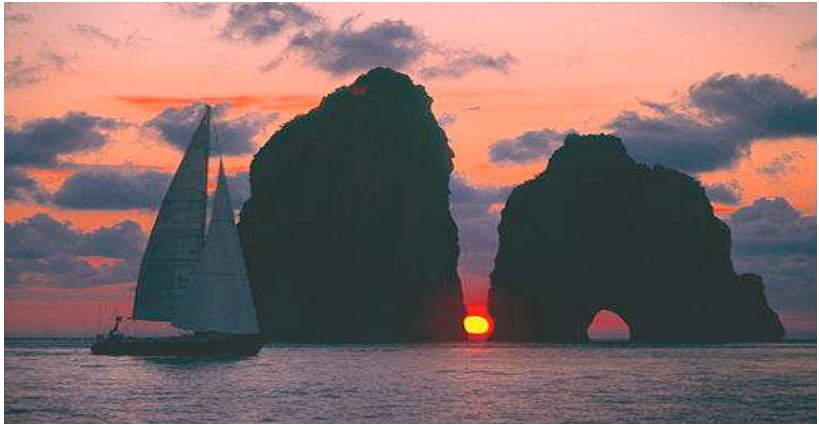}
 \caption{\it
      An allegory on HEP's future landscape. 
    \label{FUTURE} }
\end{figure}
Allow me one final comment illustrated by Fig.\ref{FUTURE}. The future landscape of 
high energy physics is dominated by two huge landmarks, namely the LHC and hopefully 
the ILC; those landmarks are represented by the two rocks on the picture. I believe there is 
still some pathway left between them for dedicated studies at a Super-Flavour facility as indicated 
by the gap between the two rocks. Heavy flavour studies thus resemble a passage between 
Scylla and Charybdis. It requires a crew and a skipper that combine experience with some 
daring to navigate through this strait -- where can we find them?

\end{document}